\documentclass[unsortedaddress,aps,preprint,showpacs]{revtex4}
\usepackage{graphicx}
\usepackage{dcolumn}
\usepackage{amsmath}
\usepackage{amssymb}
\usepackage{amsfonts}
\usepackage{bm}
\usepackage{color}
\newcommand{\be}{\begin{equation}}
\newcommand{\ee}{\end{equation}}
\newcommand{\ba}{\begin{eqnarray}}
\newcommand{\ea}{\end{eqnarray}}

\begin{document}
\title{Universal behavior of soft-core fluids\\near the threshold of thermodynamic stability}
\author{Gianpietro Malescio$^1$\footnote{Corresponding author. Email: {\tt malescio@unime.it}}, Alberto Parola$^2$\footnote{Email: {\tt alberto.parola@uninsubria.it}}, and Santi Prestipino$^1$\footnote{Email: {\tt sprestipino@unime.it}}}
\affiliation{$^1$Universit\`a degli Studi di Messina,\\Dipartimento di Scienze Matematiche e Informatiche, Scienze Fisiche e Scienze della Terra,\\viale F. Stagno d'Alcontres 31, 98166 Messina, Italy\\$^2$Universit\`a dell'Insubria,\\Dipartimento di Scienza e Alta Tecnologia,\\via Valleggio 11, 22100 Como, Italy}
\date{\today}

\begin{abstract}
We study by liquid-state theories and Monte Carlo simulation the 
behavior of systems of classical particles interacting through a finite 
pair repulsion supplemented with a longer-range attraction. Any 
such potential can be driven Ruelle-unstable by increasing the 
attraction at the expenses of repulsion, until the thermodynamic 
limit is lost. By examining several potential forms we find 
that all systems exhibit a qualitatively similar behavior in the fluid 
phase as the threshold of thermodynamic stability is approached 
(and possibly surpassed). The general feature underlying the 
approach to Ruelle instability is a pronounced widening of the 
liquid-vapor binodal (and spinodal) line for low temperatures, to 
such an extent that at the stability threshold a vanishing-density 
vapor would coexist with a diverging-density liquid. We attempt to 
rationalize the universal pathway to Ruelle instability in soft-core 
fluids by appeal to a heuristic argument.
\end{abstract}

\maketitle
\section{Introduction}

Microscopic interactions prevent atoms from overlapping each 
other, due to the strong repulsion at short distances caused by the 
Pauli exclusion principle. The situation is quite different if one 
considers interactions between macromolecules. In this case, the 
effective forces between the centers of mass, resulting from 
integrating out the internal degrees of freedom of each molecule, 
may result in a bounded repulsion. This allows such particles to 
even ``sit on top of each other'', as full overlap only costs a finite 
energy. Hence, interactions which are unphysical for atomic 
systems, may become meaningful in the context of soft matter~%
\cite{Louis1,Likos1,Likos2,Likos3}, {\it e.g.}, for polymer chains, 
dendrimers, polyelectrolytes, etc. Clearly, the
pair potential that is got after coarse graining the description
of a complex fluid will in general be both temperature- and
density-dependent; however, this does not undermine the importance
of simpler bounded potentials with fixed parameters, which remain
well suited to study the generic ({\it i.e.}, qualitative) effects of more
realistic effective interactions.

While for unbounded repulsive interactions thermodynamic stability 
is always guaranteed, since the thermodynamic limit is well defined, 
the situation is quite different for bounded repulsions. 

As first observed by Ruelle and Fisher~\cite{Ruelle,Fisher}, a pair 
potential which is bounded at the origin and enough attractive for 
some range of distances undergoes a thermodynamic catastrophe, 
{\it i.e.}, particles collapse to a finite volume of space (Ruelle 
instability). In this case, a large number $N$ of particles gather 
together into a highly dense spherical cluster, with an energy 
proportional to $N^2$. On the contrary, for thermodynamically-%
stable systems, the energy per particle is asymptotically constant 
and the system satisfies H-stability~\cite{Ruelle} ({\it i.e.}, denoting 
the total potential energy with $U$, a constant $B\ge 0$ exists such 
that $U\ge-NB$ in each system configuration). This property 
ensures that particles will not collapse as $N\rightarrow\infty$.

Ruelle and Fisher have devised a few simple criteria to check 
thermodynamic (alias H-) stability for a bounded isotropic potential 
$u(r)$, which have been recently revived by Heyes and
Rickayzen~\cite{Heyes1}. Specifically, a sufficient condition for 
Ruelle instability is $\widetilde{u}(0)<0$, $\widetilde{u}(k)$ being 
the Fourier transform of $u(r)$. Conversely, if $\widetilde{u}(k)\ge 
0$ for all $k$, then the system is thermodynamically stable. 
However, when applied to a specific bounded, parameter-%
dependent potential such criteria can only serve to locate the 
transition from the stable to the Ruelle-unstable regime, while they 
are silent on the modifications undergone by the stable system as it 
approaches the thermodynamic-stability threshold (TST).

Recently, we have studied this issue for a system of particles 
interacting through a potential consisting of a Gaussian repulsion, 
centered at the origin, augmented with a weaker Gaussian 
attraction shifted at larger distances (shifted double-Gaussian 
potential)~%
\cite{Speranza1,Speranza2,Malescio_Prestipino,Prestipino_Malescio}.
The phase behavior of this system has been investigated as a 
function of the attraction strength $\eta$. Above a certain threshold 
$\eta_{\rm c}$, the infinite-size system becomes Ruelle-unstable, 
and thus collapses to a cluster of finite volume in finite time. As
$\eta_{\rm c}$ is approached from the stable side, the liquid-vapor 
region undergoes an anomalous widening at low temperature, until 
the liquid density diverges for $T\rightarrow 0$ at $\eta=\eta_c$~%
\cite{Malescio_Prestipino,Prestipino_Malescio}. Inspired by 
previous observations by Fantoni and coworkers~%
\cite{Fantoni1,Fantoni2}, we have also analyzed the homogeneous 
fluid beyond the threshold, finding that a sharp line divides the 
thermodynamic plane in two regions, characterized by radically 
different collapsing behaviors: on one side of the line ({\it i.e.}, for 
high densities) collapse occurs extremely fast (``strongly-unstable'' 
regime), whereas on the other side (low densities) the waiting time 
for collapse enormously exceeds typical simulation time
(``weakly-unstable'' regime).

The aim of the present work is to assess whether the behavior 
observed for the shifted double-Gaussian potential holds the same 
for all systems characterized by a finite interparticle repulsion with a 
longer-range attractive component (FRAC potential). We consider a 
substantial number of FRAC potentials with widely different 
features, and examine, through the hypernetted-chain (HNC) 
equation, how their behavior changes as the attraction
becomes increasingly more effective. We find that all
systems exhibit qualitatively similar behavior below (and also beyond)
the TST. This strongly suggests that the approach of FRAC fluids
to Ruelle instability occurs along a universal pathway.
 
For all the investigated systems, the TST estimate derived from the 
HNC analysis is in excellent agreement with the value 
independently derived by the Ruelle-Fisher criteria (the discrepancy 
being of order $10^{-3}$ in relative terms, or even less). This 
supports the conclusion that, at least in a small interval around the 
threshold, the HNC predictions are reliable.

Then, for two specific FRAC interactions, namely the shifted 
double-Gaussian and the double-exponential potential, we use a 
refined liquid-state theory, the Hierarchical Reference Theory 
(HRT), as well as Monte Carlo (MC) simulation, to study how the 
shape of the liquid-vapor region changes when approaching the 
TST from below. The results confirm the previous suggestions, 
including the anomalous widening of the liquid-vapor region, thus 
indicating that the HNC theory faithfully describes the changes 
undergone by the system near the TST.

The outline of the paper is the following. In Sec.\,II we introduce a 
few parametric FRAC potentials, whose TST is exactly known from 
the Ruelle-Fisher criteria. The methods used to analyze the 
behavior close to the TST are described in Sec.\,III. In Sec.\,IV, we 
first assess the quality of a HNC-based estimate of the TST; then, 
for two specific FRAC potentials the results of the HNC analysis are 
compared with HRT and MC data. We conclude Sec.\,IV by 
providing a theoretical argument for the generic behavior of FRAC 
potentials near the TST. The last Sec.\,V is devoted to Conclusions.
 
\section{Models}
\setcounter{equation}{0}
\renewcommand{\theequation}{2.\arabic{equation}}

We here present a number of FRAC potentials which, in some 
range of parameters, become Ruelle-unstable. With one exception 
only, straightforward application of the Ruelle-Fisher criteria allows 
one to compute the TST exactly~\cite{Heyes1}. In the following, $r$ 
denotes the interparticle distance.

\vspace{2mm}
\noindent 1) {\em Double-Gaussian (DG) potential}. It consists of 
the sum of a repulsive Gaussian with an attractive one:
\be
u(r)=A\exp\{-ar^2\}-B\exp\{-br^2\}
\label{eq2-1}
\ee
(from now on, $r$ and $u(r)$ are written in dimensionless units). 
Notice that the potential form (\ref{eq2-1}) is different from the 
shifted-DG model studied in Refs.~%
\cite{Malescio_Prestipino,Prestipino_Malescio}. The DG potential is 
a generic model for the effective pair interaction between polymer 
chains in solutions. For $A>B$ and $a>b$, the DG potential has a 
positive maximum at $r=0$. As $r$ increases, $u(r)$ turns negative 
at a certain distance and, after reaching a minimum value, it 
eventually goes to zero from below when $r\rightarrow\infty$. The 
DG potential essentially depends on the ratios $A/B$ and $a/b$. 
Choosing $B=b=1$, we remain with two free parameters, $A$ and 
$a$. As $a$ increases for fixed $A$, the repulsive Gaussian 
decreases rapidly; accordingly, the attractive well of $u(r)$ 
becomes wider and deeper.

\vspace{2mm}
\noindent 2) {\em Double-exponential (DE) potential}. Its analytic 
form is:
\be
u(r)=A\exp\{-ar\}-B\exp\{-br\}
\label{eq2-2}
\ee
with $A,a,B,b\ge 0$. This potential has been employed, {\it e.g.}, in 
the modeling of small clusters~\cite{DOrsogna,Miller}. The much 
used Morse potential for neutral atoms can be written in DE form, 
with $a=2b$ and atom-specific amplitudes. For $A>B$ and $a>b$, 
the DE potential has a shape similar to the DG potential, though it 
falls more rapidly near $r=0$.

\vspace{2mm}
\noindent 3) {\em Cosine-Gaussian (CG) potential}. It is written as:
\be
u(r)=A\cos(ar)\exp\{-br^2\}
\label{eq2-3}
\ee
with $A>0$ and $a,b\ge 0$. This potential can find application in 
metals under high pressures, as an effective atom-atom interaction 
embodying the Friedel oscillations of electronic screening.

\vspace{2mm}
\noindent 4) {\em Radial symmetric short-ranged attractive 
(SHRAT) potential}. This potential, which is written as
\be
u(r)=\left\{
\begin{array}{ll}
A(1-r)^4-B(1-r)^3\,, & r\le 1 \\
0\,, & r>1\,,
\end{array}
\right.
\label{eq2-4}
\ee
has been used in the past as a generic embedded-atom potential 
for metals~\cite{Stankovic}.

\vspace{2mm}
\noindent 5) {\em Separation-shifted Lennard-Jones (LJ) potential}. 
Its form is:
\be
u(r)=Aa^{2p}(r^2+a^2)^{-p}-Bb^{2q}(r^2+b^2)^{-q}
\label{eq2-5}
\ee
with $a,b,A,B\ge 0$. At variance with the standard LJ potential, 
separation-shifted LJ potentials are finite at the origin. We will only 
consider the case $p=6$ and $q=3$. The stability of (\ref{eq2-5}) 
has been extensively studied in Ref.\,\cite{Heyes2}.

\vspace{2mm}
\noindent 6) {\em Generalized exponential$-6$ (GE6) potential}. It 
combines an exponential repulsion with an algebraic attraction 
regularized at the origin:
\be
u(r)=A\exp\{-ar\}-B(r^2+b^2)^{-3}\,.
\label{eq2-6}
\ee
If we set $b=0$ in the above expression, we obtain the exp$-6$ 
potential, also known as the Buckingham potential~%
\cite{Buckingham}, in turn a simplified case of the more general 
Born-Mayer-Huggins potential for alkali-halide crystals. The 
Buckingham potential is commonly employed as an effective pair 
potential for elemental substances under extreme thermodynamic 
conditions (see, {\it e.g.}, Refs.~\cite{Saija,Malescio}).

\section{Methods}
\setcounter{equation}{0}
\renewcommand{\theequation}{3.\arabic{equation}}

\subsection{Hypernetted-chain equation}

The phase behavior of fluids described by the potentials presented 
in Sec.\,II has been first analyzed by the HNC integral equation~%
\cite{Hansen}. In general, the HNC equation has known limitations, 
related to its thermodynamic inconsistency. However, at high 
density the HNC equation proves to be very effective in describing 
the thermodynamics and structure of particles interacting through a 
bounded pair potential (see, {\it e.g.}, Ref.~\cite{Likos4}). Hence, it 
represents a valuable tool for a systematic investigation of FRAC 
fluids.

In particular, we have looked at the boundary line (BL) separating 
the region of thermodynamic parameters where the HNC equation 
can be solved (``stable-fluid region'') from the ``unstable-fluid 
region'' where no iterative solution is found. Upon crossing the BL 
from the stable-fluid region, the computed isothermal 
compressibility $K_T$ turns abruptly from a large positive value to 
a negative one. For ordinary simple fluids, characterized by an 
unbounded short-range repulsion, the BL can roughly be 
associated with the liquid-vapor spinodal line, marking the threshold 
of instability towards phase separation (also called ``mechanical 
instability''). Whence the name ``pseudospinodal line'' also reserved 
to the BL. Clearly, no crystallization transition can be predicted by 
the HNC theory, which is a liquid-state theory, implying that a 
portion of the liquid region may actually be metastable.

Recently, the BL of the HNC equation has been computed for the 
shifted-DG model~\cite{Malescio_Prestipino,Prestipino_Malescio}. 
Compared to the liquid-vapor coexistence line obtained by 
simulation, the BL yields reasonable results. In particular, the 
binodal line and the BL show similar topological modifications as a 
function of the attraction strength.

\subsection{Hierarchical Reference Theory}

The Hierarchical Reference Theory (HRT) of fluids~\cite{Hansen,Parola}
is a genuine microscopic approach that implements renormalization-%
group considerations into a liquid-state theory. This approach 
comes closest to a realistic description of the liquid-vapor transition, 
being able to generate non-Landau critical exponents and scaling 
laws, as well as a convex free energy, so that flat isotherms at 
coexistence naturally emerge from the theory. Previous 
implementations of HRT in lattice models, atomic fluids, and mixtures 
proved its accuracy in determining the phase boundaries and the
thermodynamic properties of the systems under investigation.

In essence, HRT accounts, via a differential equation, for the effects of 
density fluctuations on top of a mean-field description of the model.
When specializing this approach to a physical system, we have first
to properly define the starting mean-field approximation. In our case,
we split the interaction $u(r)$ into the sum of a repulsive (or
``reference") and an attractive part, responsible for the occurrence of
phase separation: $u(r)=u_R(r) + w(r)$. The physical properties of the
reference system (both thermodynamics and correlations) are evaluated
via other standard liquid-state approaches, like integral equations.
In this implementation we adopted the HNC equation, which proved
accurate for systems characterized by soft-core repulsion~\cite{Bolhuis}. 
The mean-field approximation for the excess free-energy density
$A^{ex}$ then reads: $A^{ex}_{mf}=A^{ex}_R-\rho^2\widetilde{w}(0)/2$,
in terms of the number density $\rho$ of the fluid and the fully integrated
attractive part of the potential, $\widetilde{w}(0)$. 

Then, the exact HRT equation, in the sharp cut-off formulation~\cite{Parola},
describing the change in the free-energy density $A^{ex}$ upon the
inclusion of density fluctuations of wavevector $Q$ reads:
\begin{equation}
\frac{\partial A^{ex}_Q}{\partial Q}=kT\,\frac{Q^2}{4\pi^2}\ln\left[1-\rho\frac{\beta\widetilde{w}(Q)}{1-\rho\widetilde{c}_Q(Q)}\right]\,.
\label{eq3-1}
\end{equation}
This ``evolution equation" depends on the Fourier transform of the
attractive part of the interaction $\widetilde{w}(Q)$ and on the direct
correlation function $\widetilde{c}_Q(k)$ of the system when fluctuations
of wavevector larger than $Q$ have been taken into account.
The initial condition, set at $Q\rightarrow\infty$, coincides with the
mean-field approximation ($A_\infty=A_{mf}$) which disregards
fluctuations altogether, and the physical result including fluctuations
on all lengthscales is obtained for $Q=0$. However, integration of
Eq.\,(\ref{eq3-1}) requires some approximate closure, expressing the direct 
correlation function $\widetilde{c}_Q(k)$ in terms of the free-energy
density $A_Q$ at each $Q$. Most of the previous implementations of
HRT were based on a RPA-type closure, inspired by the known
Random Phase Approximation, which amounts to set
$\widetilde{c}_Q(k)=\widetilde{c}_R(k)-\beta_Q\widetilde{w}(k)$,
where the parameter $\beta_Q$ is determined by the compressibility
sum rule, valid at each $Q$:
$\frac{\partial^2 A^{ex}_Q}{\partial\rho^2}=-kT\widetilde{c}_Q(0)$. 
Within this closure, also adopted in the present study, the effects of
fluctuations on the correlations are represented as a renormalization
of the system temperature. 

The HRT equation then becomes a non-linear parabolic partial
differential equation for the free-energy density $A^{ex}$ in the
$(\rho,Q)$ plane, which has been solved numerically by use of
an implicit predictor-corrector finite-difference scheme at fixed
temperature~\cite{Parola}. The physical free energy is obtained at
$Q=0$, where the convexity requirement is always satisfied by the
theory. Below a certain temperature $T_c$ the resulting free energy
displays a region of flat isotherms, signaling the occurrence of
phase separation and allowing for the unambiguous determination
of the phase boundaries.

An illustrative example, displaying the role of
fluctuations in suppressing van der Waals loops, is shown in
Fig.\,8 of Sec.\,IV.B. The equation of state of the shifted
double-Gaussian model (defined at Eq.\,(\ref{eq4-2}) below) for
$\eta=0.025$ is plotted for different values of the cut-off $Q$, at
a reduced temperature $T=0.12<T_c$. When $Q$ is large, the
free energy is given by its mean-field value, and a van der Waals
loop is clearly present. As $Q$ is decreased, density fluctuations
of wave-vector $Q$ are gradually included, leading, for $Q=0$,
to the complete suppression of the loop: A region of constant
pressure now appears in the density interval $0.05<\rho<0.265$,
allowing the unambiguous identification of the coexistence region
at $T=0.12$.

\subsection{Monte Carlo simulation}

In Ref.\,\cite{Malescio_Prestipino}, we have investigated liquid-vapor
coexistence in a shifted-DG system near the stability threshold, using
the method of Gibbs-ensemble Monte Carlo (GEMC) 
simulation~\cite{Panagiotopoulos,Frenkel}. Here, the same analysis
is performed for a DE type of system (see Sec.\,IV.B).

In a GEMC simulation run, $N$ particles are initially distributed
between two cubic simulation boxes, and arranged in, {\it e.g.},
a lattice configuration with the same number density in both.
Periodic boundary conditions 
are applied to each box separately. One GEMC cycle consists of 
trial moves of three types: shift of a particle within a box, exchange 
of volume between the boxes, and particle swap. The acceptance 
rule as well as the schedule of the moves are designed in such a 
way that detailed balance holds; particular care has been paid in 
treating the case where one box happens to be empty
(see Ref.\,\cite{Frenkel}, which we have closely
followed in writing our GEMC code). In order to 
achieve faster equilibration at low temperature we have found 
useful to start with largely different numbers of particles in the two 
boxes (in the captions of Figs.\,8 and 9 below, we write $N_1+N_2$ 
to mean that we initially put $N_1$ particles in one box and $N_2$ 
particles in the other). Typically, from $4\times 10^4$ to $4\times 
10^5$ GEMC cycles (depending on the sample size) are more than 
enough for equilibrium quantities like the density or the energy, 
whose averages are computed over the second half of the 
trajectory only.

\section{Results}
\setcounter{equation}{0}
\renewcommand{\theequation}{4.\arabic{equation}}

\subsection{HNC results}

For each potential of those listed in Sec.\,II, we have examined first 
how the BL changes upon varying the balance between repulsive 
and attractive interactions. We focus on an interval of values of the 
potential parameters lying close to the TST.

Though the HNC equation can only provide a qualitative 
assessment of the fluid phase diagram, it turns out that it is 
extremely accurate in locating the ultimate threshold of 
thermodynamic stability. As shown below, for all the investigated 
systems we find that the HNC estimate of the TST is in excellent 
agreement with the analytic result derived from the Ruelle-Fisher 
criteria (relative differences being of the order of $10^{-3}$, or even 
smaller). Therefore, we can confidently assume that, at least in a 
small interval around the threshold, the HNC equation is able to 
grasp the essential features of the fluid behavior. At the end of the 
section, we present a theoretical argument to justify the 
effectiveness of the HNC equation for FRAC fluids.

\vspace{2mm}
\noindent 1) {\em DG potential}. For fixed $A,B,b$, the DG potential 
(\ref{eq2-1}) only depends on the $a$ parameter. As $a$ increases, 
the repulsion becomes more and more short-ranged, hence the
interval of distances over which the attractive component
is effective grows, and the depth of the attractive well also increases.
Clearly, the attraction can be enhanced relative to repulsion in many
different ways; for example, at fixed $A,a,b$, an increase of $B$
makes the attractive well deeper, while the attractive range is not
affected. Otherwise, one can extend the range of the attractive well
(controlled by $b$) while keeping its depth ($B$) fixed. In general,
for any given potential one can use different parameters to control
the importance of attraction versus repulsion. The choice of the
control parameter determines the way in which attraction grows
at the expense of repulsion as the system approaches thermodynamic
instability. Alternatively, one may consider many different potential forms
and study their behavior in the approach to thermodynamic instability
as only one of the possible control parameters is varied. In our analysis
we followed the latter option, which allows a more thorough
investigation of the space of interaction potentials.

%
%
\begin{figure}[t]
\includegraphics[width=10cm]{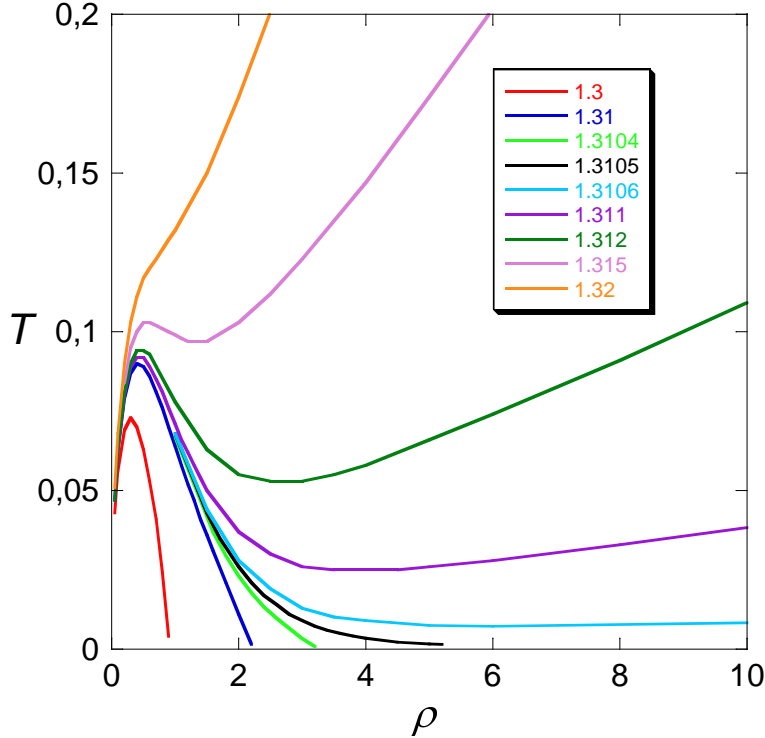}
\caption{DG potential for $A=1.5,B=b=1$: BL for a few values of
$a$ (in the legend), growing from bottom to top. In this picture, as 
well as in the following ones, both $\rho$ and $T$ are expressed in 
dimensionless units.}
\label{fig1}
\end{figure}

%
%
\begin{figure}[t]
\includegraphics[width=10cm]{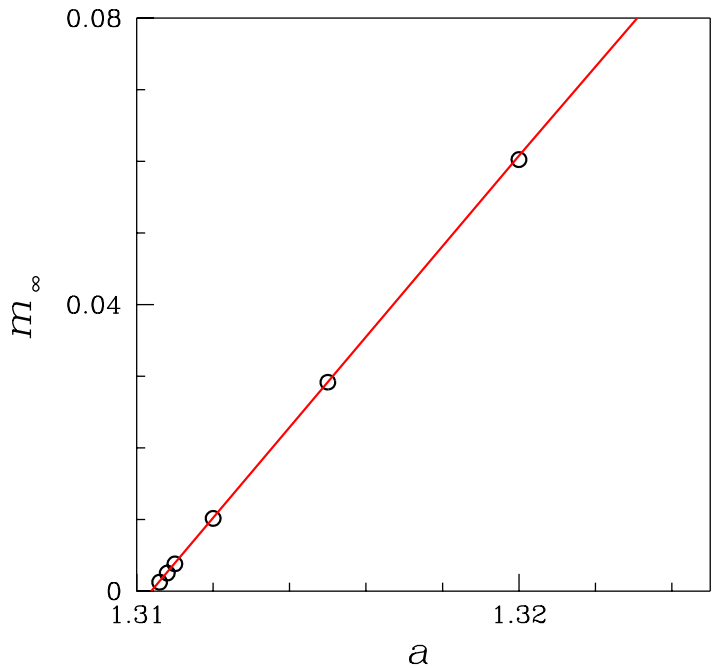}
\caption{DG potential for $A=1.5,B=b=1$: asymptotic BL slope
$m_\infty$ for a number of $a$ values (open dots). A linear least-%
square fit of the data points gives $a'_c=1.31040247$, to be 
compared with $a_c=1.31037069\ldots$ The red line represents
$|\widetilde{u}(0)|$ as a function of $a$ (see Sec.\,IV.C). The $r$ 
grid used in the computer code to solve the HNC equation has 
2048 points and a spacing of 0.01 (see text).}
\label{fig2}
\end{figure}

In Fig.\,1 we plot the BL for $A=1.5$ and $B=b=1$ in a range of
$a$ values enclosing the TST ($a=a_c=1.31037069\ldots$). We 
see in this picture that, for small $a$, the unstable-fluid region is 
bell-shaped and its area is finite; above the temperature
$T_{\rm max}$, corresponding to the bell maximum, the HNC 
equation can be solved for any $\rho$. As $a$ increases, the 
unstable-fluid region becomes higher ($T_{\rm max}$ increases) 
and wider. In the proximity of $a_c$, the widening of the curve 
blows up at low $T$, until the liquid density apparently diverges for 
$T\rightarrow 0$ at $a=a'_c\simeq a_c$. In analogy to the evolution 
of the liquid-vapor binodal line in the shifted-DG model~%
\cite{Malescio_Prestipino}, we surmise that $a'_c$ is the HNC 
estimate of $a_c$. For $a>a'_c$ the topology of the BL changes 
radically: the right side of the bell ``opens up'' and, at large 
densities, the BL becomes a straight line with positive slope. In 
other words, for all $T$ there is a density above which the HNC 
equation cannot be solved. Accordingly, the area of the unstable-%
fluid region becomes infinite. As $a$ increases further, the BL 
becomes monotonically increasing, and eventually resembles a 
straight line running close to the $T$ axis; in turn, the unstable-fluid 
region covers the entire thermodynamic plane, except only for a 
narrow region of low densities.

In order to estimate $a'_c$ accurately, we proceed as follows. At 
$a=a'_c$, the BL has a vanishing asymptotic slope. We calculate 
for $a>a'_c$ the slope $m_\infty$ of the BL at very high density 
(where the BL is a perfectly straight line) and report it in a graph as 
a function of $a$ (Fig.\,2). We find that these slopes lie on a straight 
line, meaning that the BL slope is to a very good approximation a 
linear function of $a$, at least close enough to $a'_c$. Through a 
least-square fit, we extrapolate $m_\infty(a)$ down to zero slope, 
thus obtaining the crossover value $a'_c$ between the stable and 
the Ruelle-unstable regime. We stress that $m_\infty(a)$ is close to 
a straight line for all the $a>a'_c$ values considered in Fig.\,1, 
regardless of the shape of the BL at low density.

Obviously, the accuracy of $a'_c$ depends on the $r$ grid used to 
solve the HNC equation by iteration. Throughout this paper, we use 
a spatial grid with $N=2^{10}$ points and spacing $0.02$ (we have 
performed a few checks with more refined grids and found no 
appreciable variation in the overall BL behavior). We obtain 
$a'_c=1.3105168$, which is extremely close to $a_c$ (the 
difference is 0.00015). Using a denser mesh of points ($N=2^{11}$, 
with spacing $0.01$) we obtain the even more accurate estimate 
$a'_c=1.31040247$ (corresponding to a difference $a'_c-a_c$ of 
0.000032). A similar systematic improvement in accuracy is found 
for all the potentials considered.

%
%
\begin{table}[t]
\caption{DG potential for $B=b=1$: For a few $A$ values we 
compare the TST $a_c$ with its estimate $a'_c$ from the HNC 
theory. The $a'_c$ data have been computed with a $r$ grid of 
1024 points and a spacing of 0.02 (see text).}
\begin{center}
\begin{tabular}{ccc}
\hline\hline
$A$ & $a_c$ & $a'_c$\\
\hline
1.5 & 1.3104 & 1.31052(1) \\
2 & 1.5874 & 1.5875(5) \\
3 & 2.0801 & 2.0805(5) \\
\hline\hline
\end{tabular}
\end{center}
\end{table}

In Table I we report $a'_c$ for a few values of $A$, and compare it 
to the TST (according to the Ruelle-Fisher criteria, $A/B\ge
(a/b)^{3/2}\ge 1$ is a necessary and sufficient condition for 
thermodynamic stability~\cite{Heyes1}). It turns out that $a'_c$ is in 
remarkable agreement with the threshold $a_c$, the relative 
discrepancy being smaller than $0.001$.

%
%
\begin{figure}[t]
\includegraphics[width=10cm]{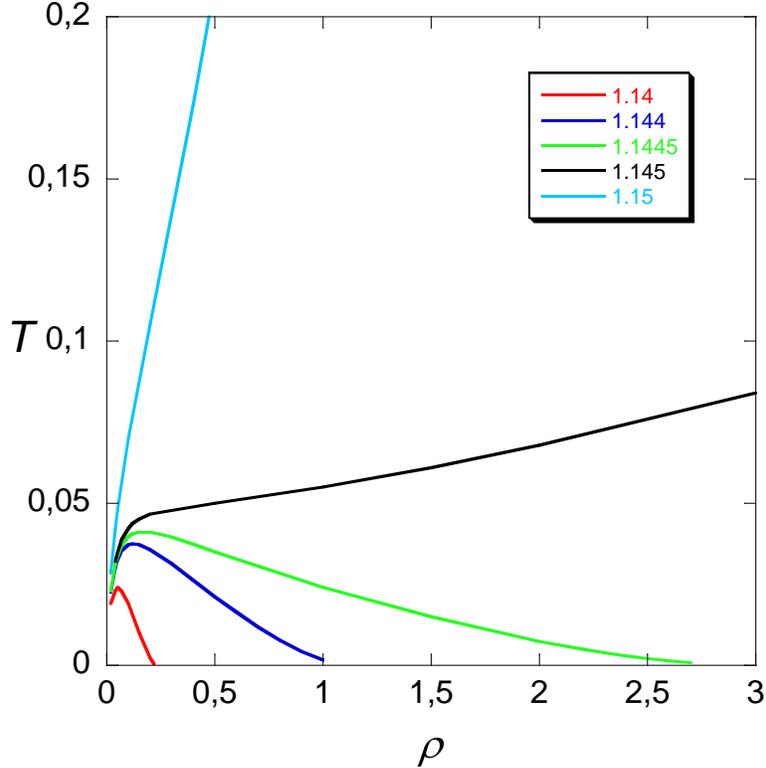}
\caption{DE potential for $A=1.5,B=b=1$: BL for a few values of
$a$ (in the legend), growing from bottom to top.}
\label{fig3}
\end{figure}

%
%
\begin{table}[t]
\caption{DE potential for $B=b=1$: For a few $A$ values we 
compare the TST $a_c$ with its estimate $a'_c$ from the HNC 
theory.}
\begin{center}
\begin{tabular}{ccc}
\hline\hline
$A$ & $a_c$ & $a'_c$\\
\hline
1.5 & 1.1447 & 1.1447(3) \\
2 & 1.2599 & 1.2595(5) \\
3 & 1.4422 & 1.4425(5) \\
\hline\hline
\end{tabular}
\end{center}
\end{table}

\vspace{2mm}
\noindent 2) {\em DE potential}. For fixed $A,B,b$, the DE potential 
(\ref{eq2-2}) only depends on $a$. As $a$ increases, the repulsion 
becomes more short-ranged and the importance of the attraction in 
turn increases. The evolution of the BL with increasing $a$ (see 
Fig.\,3) is similar to that reported for the DG potential.

In Table II we report $a'_c$ for a few values of $A$, while $a_c$ 
follows from the Ruelle-Fisher criteria, according to which
$A/B>(a/b)^3$ is a necessary and sufficient condition for 
thermodynamic stability~\cite{Heyes1}. The relative discrepancy 
between the HNC threshold and the exact one is smaller than 
0.001.

\vspace{2mm}
\noindent 3) {\em CG potential}. According to the Ruelle-Fisher 
criteria, $a^2/(2b)\le 1$ is a necessary and sufficient condition for 
thermodynamic stability~\cite{Heyes1}. A CG-type of potential has 
been considered by Louis {\em et al.}~\cite{Louis2}, who write it in 
the form:
\be
u(r)=\cos\left(\sqrt{2+\delta}\,r\right)\exp\{-r^2\}
\label{eq4-1}
\ee
({\it i.e.}, $a=\sqrt{2+\delta},b=1$). Hence, the stability criterion for 
(\ref{eq4-1}) reads $\delta\le 0$. Our analysis based on the HNC 
equation yields a threshold value $\delta'_c=0.001(1)$ (see Fig.\,4). 

%
%
\begin{figure}
\includegraphics[width=10cm]{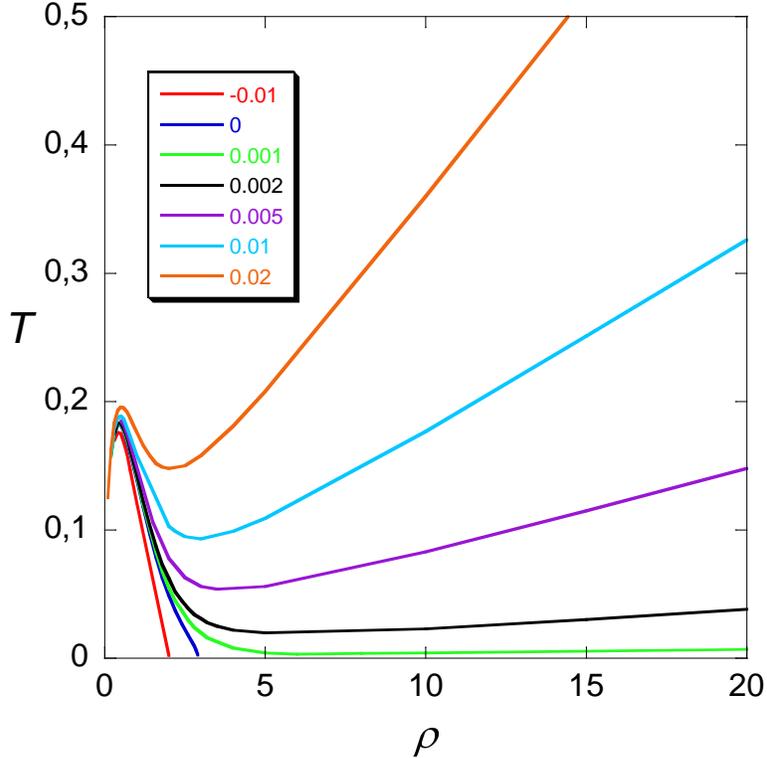}
\caption{CG potential (\ref{eq4-1}): BL for a few values of $\delta$ 
(in the legend), growing from bottom to top.}
\label{fig4}
\end{figure}

\vspace{2mm}
\noindent 4) {\em SHRAT potential}. According to the Ruelle-Fisher 
criteria, $A/B\ge 7/4$ is a necessary and sufficient condition for 
thermodynamic stability~\cite{Heyes1}. For $B=1$, the HNC 
analysis gives $A'_c=1.7497(5)$ (see Fig.\,5). Notice that the path 
from the stable to the Ruelle-unstable regime corresponds for the 
SHRAT potential to the direction of decreasing $A$. 

%
%
\begin{figure}
\includegraphics[width=10cm]{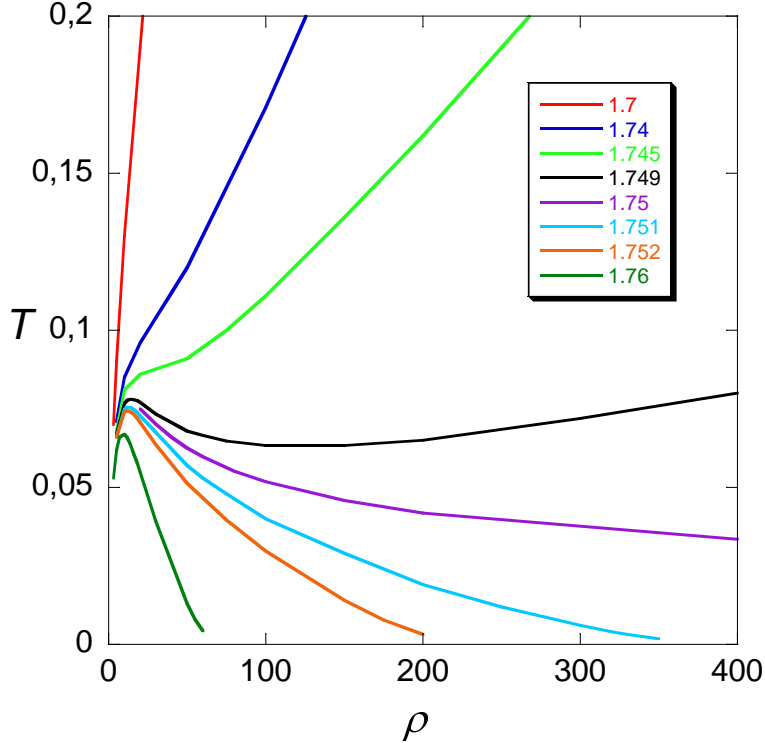}
\caption{SHRAT potential for $B=1$: BL for a few values of $A$ (in 
the legend), growing from top to bottom.}
\label{fig5}
\end{figure}

\vspace{2mm}
\noindent 5) {\em Separation-shifted LJ potential}. In
Ref.\,\cite{Heyes2}, the conditions for stability of potential 
(\ref{eq2-5}) have been investigated for $p=6$ and $q=3$ (which 
yields a regularized LJ potential), and it has been found for $a=b$ 
that $A/B\ge 32/7=4.5714\ldots$ is a necessary and sufficient 
condition for thermodynamic stability. For $B=1$, the HNC equation 
predicts a threshold value $A'_c=4.569(1)$ (see Fig.\,6). Again, the 
system turns from stable to Ruelle-unstable upon decreasing the 
value of $A$.

%
%
\begin{figure}
\includegraphics[width=10cm]{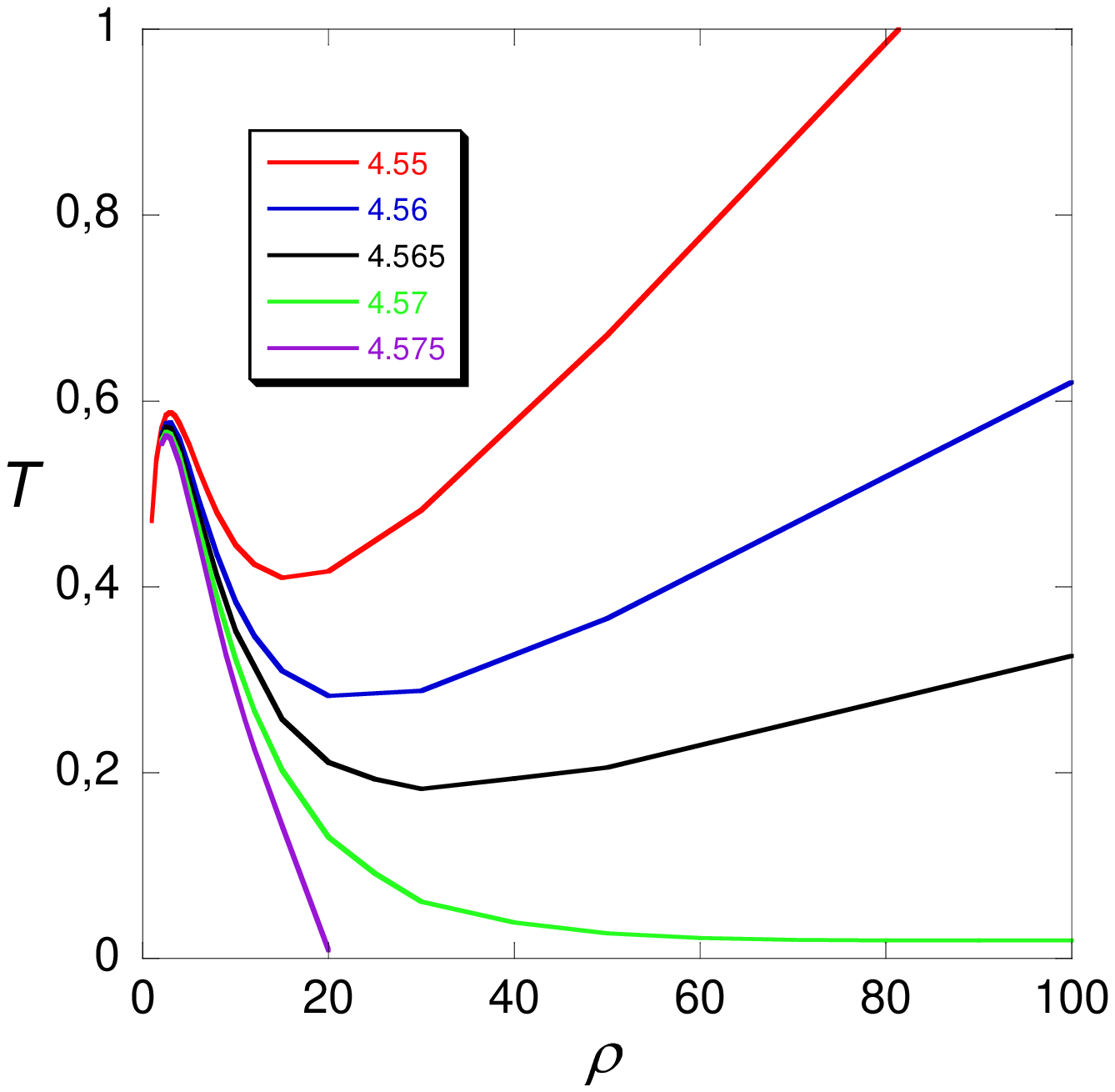}
\caption{Separation-shifted LJ potential ($a=b,B=1,p=6,q=3$): BL 
for a few values of $A$ (in the legend), growing from top to bottom.}
\label{fig6}
\end{figure}

\vspace{2mm}
\noindent 6) {\em GE6 potential}. The Fourier transform of the GE6 
potential in Eq.\,(\ref{eq2-6}) is negative at the origin for
$Ab^6/B<\pi (ab)^3/32$~\cite{Heyes1}. However, $\widetilde{u}(k)$ 
is not positive definite for $Ab^6/B\ge\pi (ab)^3/32$, hence one 
cannot state --- based only on the Ruelle-Fisher criteria --- that the 
system is thermodynamically stable above this threshold. However, 
the HNC analysis indicates that this is likely the case. We have 
computed the BL for $B=b=1,a=3$, and several $A$ values (see 
Fig.\,7; notice that the system turns from stable to Ruelle-unstable 
with decreasing $A$). By fitting the asymptotic BL slopes for 
$A=2.6,2.62,2.64$, we obtain $A'_c=2.6503(2)$, close indeed to 
the TST value ($A_c=2.6507$). We have checked numerically that, 
for $A=2.65$, the BL indeed attains a minimum for a density
$\rho\approx 200$.  

%
%
\begin{figure}
\includegraphics[width=10cm]{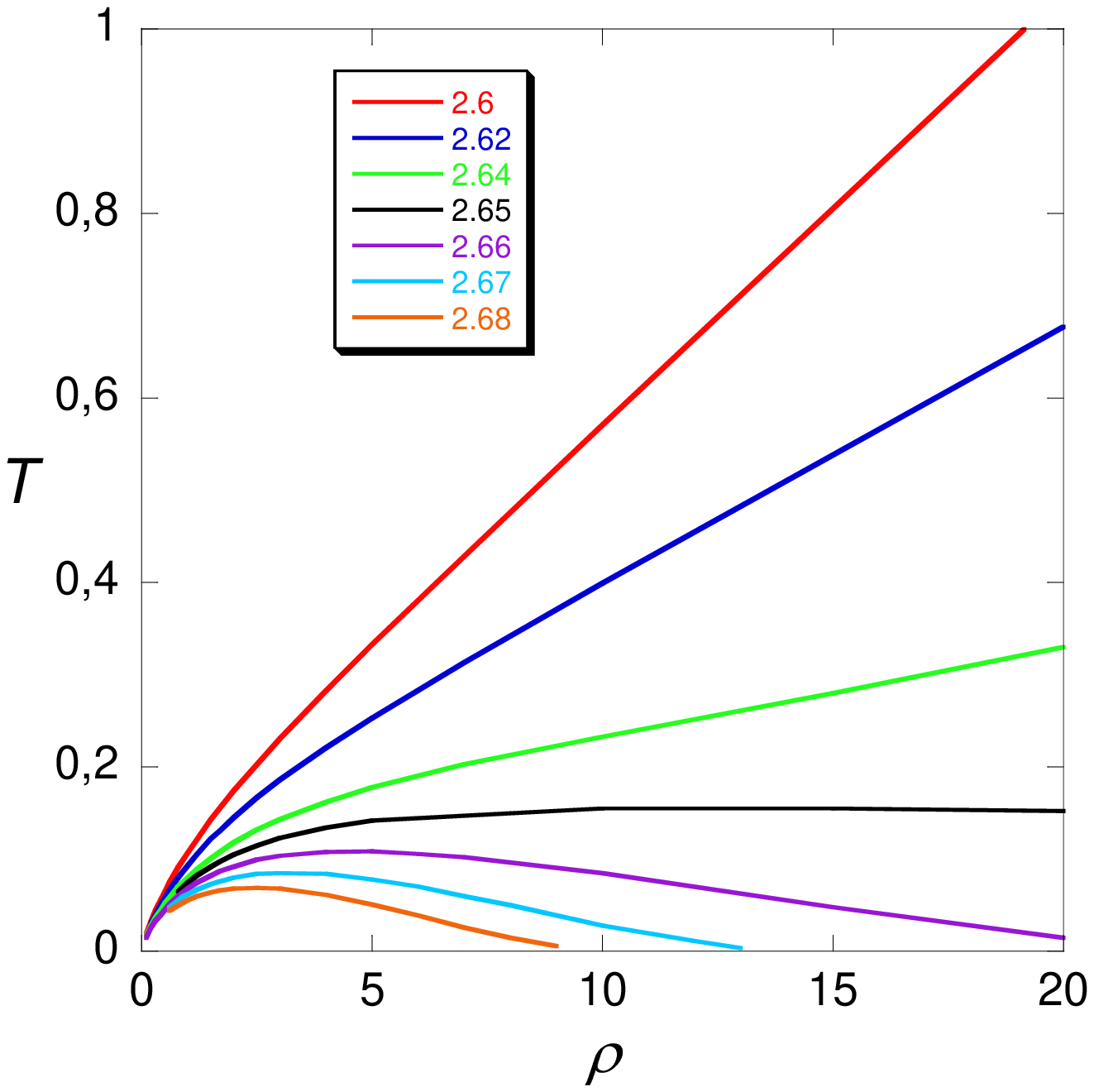}
\caption{GE6 potential for $B=b=1,a=3$: BL for a few values of $A$ 
(in the legend), growing from top to bottom.}
\label{fig7}
\end{figure}

\subsection{Comparison with HRT and MC results}

%
%
\begin{figure}
\includegraphics[width=12cm]{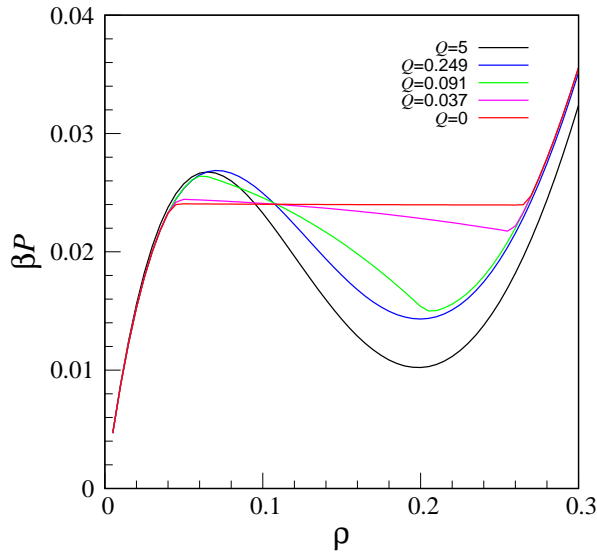}
\caption{The HRT equation of state (reduced pressure
vs. density) of the model defined in Eq.\,(\ref{eq4-2}) for $\eta=0.025$
at different values of the cut-off $Q$. The initial condition at $Q=5$
is given by the mean-field approximation (black line). Integration of
the ``evolution equation'' (\ref{eq3-1}) allows to gradually include
density fluctuations. The equation of state is shown for a few
representative intermediate values of the cut-off: $Q=0.249$ (blue),
$Q=0.091$ (green), and $Q=0.037$ (magenta), till $Q=0$ (red),
where convexity of free energy is eventually recovered.}
\label{fig8}
\end{figure}

%
%
\begin{figure}[t]
\centering
\includegraphics[width=10cm]{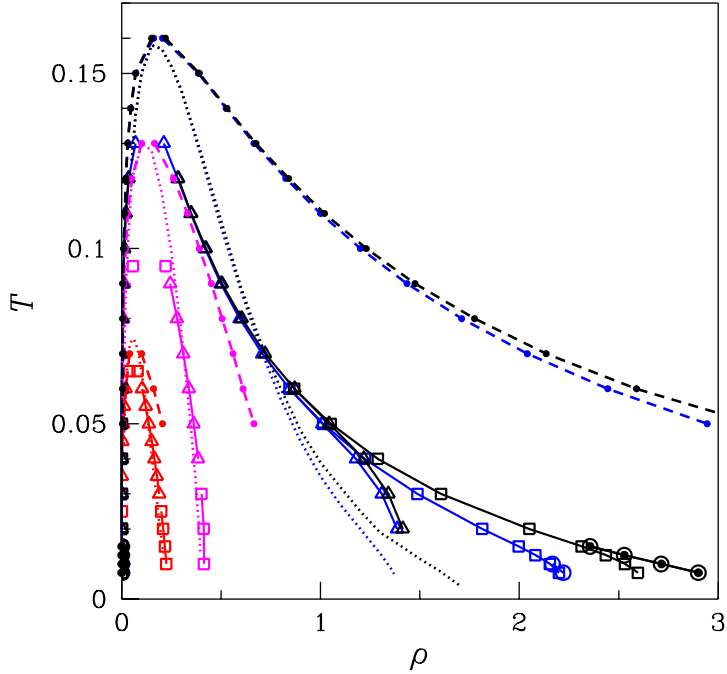}
\caption{Shifted-DG potential (see Eq.\,(1) of
Ref.\,\cite{Malescio_Prestipino}) below
$\eta_{\rm c}=0.02631579\ldots$: Liquid-vapor coexistence 
densities from GEMC simulations performed for a number of $\eta$ 
values (0.020, red; 0.025, magenta; 0.0263, blue; and 0.02631, 
black). The various symbols correspond to different initial numbers 
of particles in the two simulation boxes ($864+864$, triangles; 
$4000+108$, squares; $8192+128$, open dots; and $16000+128$, 
full dots). Also reported are the BL and the HRT coexistence 
densities for the same $\eta$ values (dotted and dashed lines, 
respectively).}
\label{fig9}
\end{figure}

%
%
\begin{figure}[t]
\centering
\includegraphics[width=10cm]{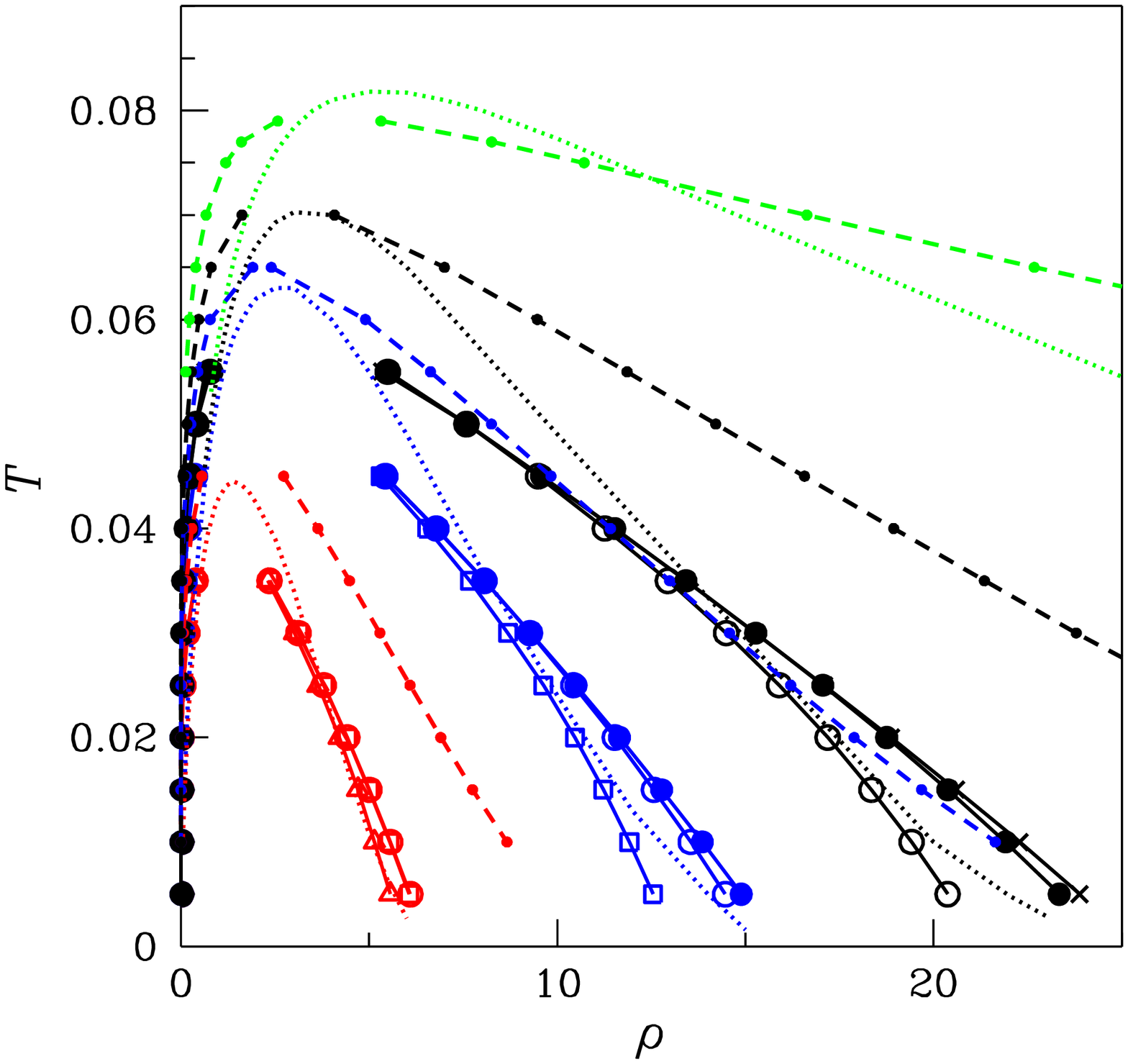}
\caption{DE potential (\ref{eq4-2}) above $L_c=0.79370052\ldots$: 
Liquid-vapor coexistence densities from GEMC simulations 
performed for a number of $L$ values (0.80, red; 0.796, blue; and 
0.795, black). The various symbols correspond to different initial 
numbers of particles in the two simulation boxes ($864+864$, 
triangles; $4000+256$, squares; $8192+250$, open dots; 
$16000+128$, full dots; and $32000+256$, crosses). Also reported 
are the BL and HRT coexistence densities as dotted and  dashed 
lines, respectively (the green lines are for $L=0.794$).}
\label{fig10}
\end{figure}

For the same shifted-DG potential investigated in
Ref.\,\cite{Malescio_Prestipino}, {\it i.e.},
\be
u(r)=\exp\{-r^2\}-\eta\exp\{-(r-3)^2\}\,,
\label{eq4-2}
\ee
we have used the HRT to compute a number of points along the 
binodal line for a few values of $\eta$ (the attraction strength). 
Adding these points to the MC data points in Fig.\,1 of
Ref.\,\cite{Malescio_Prestipino}, we finally obtain the present
Fig.\,9. As $\eta$ moves towards $\eta_c$, the HRT binodal line 
shows the characteristic widening already observed in the BL and 
in the MC data. Moreover, the HRT binodal line satisfactorily 
encloses the HNC pseudospinodal line. In more quantitative terms, 
however, the HRT substantially overestimates the densities of the 
coexisting liquid.

As a second example, we have considered the following one-%
parameter family of DE potentials,
\be
u(r)=2\exp\{-3r/L\}-\exp\{-3r\}\,.
\label{eq4-3}
\ee
The TST of potential (\ref{eq4-3}) is $L_c=2^{-1/3}$ (see
Sec.\,IV.A). In order to speed up the simulation, the potential has 
been cut off at $r=5$. We have carried out GEMC simulations on 
samples of increasing size, until finite-size effects vanish altogether. 
We have checked that, once formed, the liquid always occupies the 
box of larger volume (otherwise, the simulation is restarted from a 
different configuration). The BL, the MC coexistence line for various 
sizes, and the HRT binodal line are all reported in Fig.\,10. Overall, 
the same scenario of Fig.\,9 shows up, with the HRT line 
considerably wider than the MC line, especially closer to $L_c$. 
Despite this, the behavior of the HRT line qualitatively reproduces 
the MC coexistence line.

As a last comment, we point out that we have never observed in 
our simulations the spontaneous onset of a crystalline phase. This 
means that, either a stable crystal only occurs at densities 
considerably higher than those probed here, or the liquid phase is 
actually metastable but its lifetime is very long. In any event, the 
possible metastability of the liquid phase at the highest densities 
would not weaken in any respect the expectation, corroborated by 
simulation in two distinct cases, of an anomalous widening of the 
liquid-vapor coexistence line of FRAC fluids approaching the TST 
from below.

\subsection{Theory}

In order to account for the BL behavior described in Sec.\,IV.A, and 
thus unveil the roots of universality in the approach of FRAC fluids 
to the TST, we put forward an argument that builds on that 
presented in Ref.\,\cite{Malescio_Prestipino} but considerably 
extend it. In the following, the ultimate threshold of stability of the 
homogeneous fluid is identified with the locus $T_0(\rho)$ where 
the inverse isothermal compressibility, given in general by
\be
\frac{1}{\rho k_BTK_T}=\frac{1}{1+\rho\widetilde{h}(0)}=1-\rho\widetilde{c}(0)\,,
\label{eq4-4}
\ee
is predicted to vanish.

In Ref.\,\cite{Malescio_Prestipino}, we have made the hypothesis 
that, for each FRAC fluid, a characteristic density $\rho_0(T)$ 
exists above which the total correlation function
\be
h(r)\approx 0\,,
\label{eq4-5}
\ee
that is, $h$ is zero for all distances to within a small tolerance fixed 
once and for all (we have checked that Eq.\,(\ref{eq4-5}) is roughly 
satisfied in MC simulations of the shifted-DG fluid along the
$\rho=5$ isochore). We expect that $\rho_0(T)$ is a decreasing 
function of $T$, because the higher is the temperature the better 
the system would conform to ideal-gas behavior. Equation 
(\ref{eq4-5}) tells that the structure of a FRAC fluid at high enough 
density resembles that of an ideal gas (``infinite-density ideal gas'' 
limit~\cite{Stillinger,Likos1}).

This ansatz is actually a direct consequence of the form of the
HNC equation:
\be
c(r)=h(r)-\ln[1+h(r)]-\beta u(r)
\label{eq4-6}
\ee
In fact, by spatial integration of both sides of Eq.\,(\ref{eq4-6}),
we get 
$\widetilde{c}(0)=\int{\rm d}{\bf r}\,\left\{h(r)-\ln[h(r)+1]\right\}-\beta\widetilde{u}(0)$.
The HNC equation, by encompassing the Ornstein-Zernike equation,
has solutions only provided that $\widetilde{c}(0)<1/\rho$,
which implies
$\int{\rm d}{\bf r}\,\left\{h(r)-\ln[h(r)+1]\right\}<\beta\widetilde{u}(0)+1/\rho$.  
Close to the Fisher-Ruelle stability threshold and at high density,
the right-hand side is small and thermodynamic stability then
implies that also the left-hand side is small. However, the integrand
is positive semidefinite and vanishes only for $h(r)=0$. As a consequence,
in this limit the HNC equation forces the total correlation function
to small values at all distances.

If Eq.\,(\ref{eq4-5}) holds, for $\rho>\rho_0(T)$ we obtain from the 
HNC equation that
\be
c(r)=h(r)-\ln[1+h(r)]-\beta u(r)\approx -\beta u(r)\,,
\label{eq4-7}
\ee
namely, at sufficiently high density the HNC approximation reduces 
to the random phase approximation (RPA). Taken the RPA for 
granted, we have:
\be
\frac{1}{\rho k_BTK_T}=1+\beta\widetilde{u}(0)\rho\,.
\label{eq4-8}
\ee
Hence, $K_T$ is always positive in the stable regime,
where $\widetilde{u}(0)>0$, while it is negative beyond the density
$k_BT/|\widetilde{u}(0)|$ in the unstable regime (for the sake of clarity,
we hereafter use a notation appropriate to the shifted-DG potential~%
\cite{Malescio_Prestipino,Prestipino_Malescio}, where the stable
regime corresponds to $\eta<\eta_c$; however, similar considerations
apply for each parametric FRAC potential which becomes
Ruelle-unstable exactly where $\widetilde{u}(0)$ changes sign). 
While the previous conclusions are consistent with the high-density 
behavior of the BL for $\eta>\eta_c$~\cite{Malescio_Prestipino}, 
they are clearly insufficient to explain its anomalous widening below 
$\eta_c$.

Reasoning in purely heuristic terms, a better approximation for 
large densities would be:
\be
\widetilde{c}(0)=-\beta\widetilde{u}(0)+\frac{A}{\rho}\,,
\label{eq4-9}
\ee
$A$ being a dimensionless quantity. From Eq.\,(\ref{eq4-4}), it 
then follows:
\be
\frac{1}{\rho k_BTK_T}=1+\beta\widetilde{u}(0)\rho-A\,.
\label{eq4-10}
\ee
Aiming to reproduce the HNC phenomenology at high density, we 
assume that $A=T^*/T$, for a convenient $T^*$. This is a 
reasonable assumption, considering that Eq.\,(\ref{eq4-6}) 
becomes better satisfied, at fixed density, when $T$ is higher. With 
this $A$, we obtain:
\be
K_T>0\,\,\,\Longleftrightarrow\,\,\,k_BT>k_BT^*-\widetilde{u}(0)\rho\,.
\label{eq4-11}
\ee
Below $\eta_c$, Eq.\,(\ref{eq4-11}) predicts that the BL is a straight 
line for high densities. Stated differently, Eq.\,(\ref{eq4-11}) reads:
\be
K_T>0\,\,\,\Longleftrightarrow\,\,\,\rho>\frac{k_B(T^*-T)}{\widetilde{u}(0)}\,.
\label{eq4-12}
\ee
In particular, at $T=0$ the system would be stable only for densities 
larger than
\be
\rho^*=\frac{k_BT^*}{\widetilde{u}(0)}\,,
\label{eq4-13}
\ee
which grows and eventually diverges when $\eta$ goes to $\eta_c$. 
Very close to $\eta_c$, where
$\widetilde{u}(0)\approx B(\eta_c-\eta)$ with $B>0$, an immediate 
prediction is $\rho^*\propto(\eta_c-\eta)^{-1}$. However, we find that 
this scaling law is not generally obeyed, which can only be the 
effect of a $\rho_0(T)$ diverging for $T\rightarrow 0$ faster than
$\rho^*$. Indeed, we have checked in a few cases that the radial 
structure of the saturated liquid near $T=0$ is anything but trivial in 
the HNC theory. We conclude that, even though the simple 
modification (\ref{eq4-9}) to the RPA correctly accounts for the 
general blowing up of $\rho^*$ at $\eta_c$, it is by far insufficient to 
give the correct scaling exponent (which, moreover, appears to be 
non-universal).

{\em Above} $\eta_c$, where the HNC equation can still be solved, 
we obtain from Eq.\,(\ref{eq4-11}):
\be
K_T>0\,\,\,\Longleftrightarrow\,\,\,k_BT>k_BT^*+|\widetilde{u}(0)|\rho\,.
\label{eq4-14}
\ee
In particular, the asymptotic slope of the BL above $\eta_c$ is just 
$|\widetilde{u}(0)|$, as already checked to a very high precision in 
the shifted-DG potential~\cite{Malescio_Prestipino} (see also
Fig.\,2). Stated differently, the stability condition reads
\be
\rho<\frac{k_B(T-T^*)}{|\widetilde{u}(0)|}\,,
\label{eq4-15}
\ee
and is clearly violated for $T=0$, with the result that the
unstable-fluid region now extends to infinite density.

\section{Conclusions}

We have shown that a considerable number of FRAC potentials 
exhibit similar fluid behavior when the threshold of thermodynamic 
(alias H-) stability is approached, and eventually surpassed. The 
most important feature of this general behavior is the pronounced 
widening of the binodal line at low $T$. Right at the TST, a 
vanishing-density vapor coexists with a diverging-density liquid. 
This is consistent with the long-time behavior of the
shifted double-Gaussian fluid beyond the
TST~\cite{Malescio_Prestipino,Prestipino_Malescio},
where $N$ particles collapse to an extremely dense 
aggregate and the energy per particle is proportional to
$N$. Given the 
widely different features of the systems examined in
this paper, our results 
suggest that the transition of FRAC fluids to Ruelle instability 
indeed occurs following a universal pathway.
 
Using the HNC pseudospinodal line as a clue to liquid-vapor 
coexistence behavior, we have found that the HNC estimate of the 
TST is, for all the investigated systems, in excellent agreement with 
the value provided by the Ruelle-Fisher criteria. This supports the 
assumption that, at least in a narrow interval around this threshold, 
the HNC equation gives reliable indications. For two specific 
interactions (the shifted-DG and DE potentials), the predictions of 
the HNC equation have been checked against MC simulation and a 
more refined liquid-state theory, the HRT. By this comparison, we 
conclude that the HNC equation faithfully describes the 
modifications undergone by the liquid-vapor coexistence line as the 
TST is approached from the stable side. We argue that this happens
as a result of the nearly ideal-gas structure of the high-density
fluid, as illustrated in Sec.\,IV.C, where we have put
forward a general explanation for the universal approach of a
bounded potential to the TST, by an argument that supersedes and
improves the original one given in Ref.\,\cite{Malescio_Prestipino}.

In the light of the present results, it is possible to build up the 
following general scenario for the transition to Ruelle instability in 
fluid systems with a bounded interparticle repulsion and a longer-%
range attraction. For stable homogeneous fluids, there exists a 
region of mechanical instability (bounded above by the spinodal 
line) lying inside that of thermodynamic instability (bounded above 
by the binodal line). Both regions occupy a bounded subset of the 
thermodynamic plane. Upon getting closer to the threshold of 
Ruelle instability, the extension of both regions increases due to 
widening of both binodal and spinodal lines at low temperatures. 
When the TST is eventually reached, the liquid density becomes 
infinitely large for $T\rightarrow 0$. In the HNC analysis, the 
transition to the Ruelle-unstable regime is evidenced in the 
``opening up'' of both thermodynamic and mechanical unstable-fluid 
regions, {\it i.e.}, the extension of these regions on the $\rho$-$T$
plane turns from bounded to unbounded at the crossing of the TST.

\end{document}